\documentstyle[aps,twocolumn]{revtex}
\input epsf
\begin{document}

\wideabs{\title{Diffusion Thermopower of Ferromagnetic Transition metals}
\author{P.K. Thiruvikraman}
\address{Raman Research Institute, Bangalore-560 080, India}
\maketitle
\begin{abstract}

This paper discusses a simple calculation for the diffusion thermopower of a transition metal ferromagnet. The main
result of this calculation is that unlike in the case of a free-electron model, the diffusion thermopower $S$ and
its derivative with respect to temperature $dS/dT$ have opposite signs. The results of this calculation agree
qualitatively with experimental results in the high temperature region, where the diffusion thermopower is dominant.

 \end{abstract}}

\section{Introduction}
		Thermoelectric power (TEP) is one of the most well studied transport properties in metals
\cite{Barnard,ssp series}. It has
been
shown that it is more sensitive than resistivity to changes in the electronic structure of a material . TEP
has one
more advantage in that, usually, the sign of the thermopower depends on the sign of the charge carriers as it
depends on the electronic charge e, while resistivity is proportional to the square of the charge and hence it
has no information on the sign of the charge carriers. At the same
time, interpretation of TEP data is complicated by this very sensitivity to the various parameters involved.

		The thermoelectric power of a material is usually separated into the diffusion thermopower and the
contribution due to 'phonon drag' \cite{Barnard}. In case of a magnetic material, there are additional contributions
due to 'magnon
drag'. It has been shown that the contributions due to phonon and magnon drag are significant only at low
temperatures \cite{Barnard}. Hence for temperatures higher than the Debye
temperature the calculation of the thermopower can be
restricted to the diffusion component.

		The diffusion thermopower is supposed to be a linear function of temperature, according to the free
electron theory of metals \cite{Barnard}. In contrast to this, the thermopower of most transition metals is highly
non-linear,
indicating the inapplicability of the free-electron model to the transition metals. 

\section{Diffusion Thermopower}
The TEP results are usually interpreted in terms of the Mott's formula for diffusion thermopower \cite{Barnard}:

\begin{equation}
S_{diff} = \left [\frac{\pi ^2k^2T}{3e}\frac{dln\sigma}{dE}\right ]_{E = E_F} \label{eq:tep}
\end{equation}

Where $\sigma$ is the conductivity and $e$ is the electronic charge. Therefore the thermopower is given by the
derivative of the electrical conductivity with respect to energy, evaluated at the Fermi energy.

The electrical conductivity is given by Drude's formula:
\begin{equation}
\sigma = \frac{ne^2\tau}{m}
\end{equation}

This equation is based on the free-electron approximation.
For transition metals, it is more appropriate to write this formula in the following form:

\begin{equation}
\sigma(E_F) = \frac{2}{3}e^2v_s^2\tau (N_s(E))_{E = E_F} \label{eq:sigma}
\end{equation}

Where $\tau$ is the relaxation time and $N_s$ is the density of states of s-electrons. In
the case of
transition metals the predominant mechanism for scattering is
the scattering of the s electrons responsible for conduction into the less-mobile d states. The relaxation time will
be less, if there are more number of d-states available for the scattered electrons.  Therefore
\begin{equation}
\tau=\frac{1}{N_d} \label{eq:tau}
\end{equation}

Where $N_d$ is the density of d-states.

Thus,
\begin{equation}
\sigma(E_F) = Av_s^2\left [\frac{N_s(E)}{N_d(E)}\right ]_{E = E_F}
\end{equation}

where $A$ is a constant with respect to energy

In the case of transition metals like Iron it is well known that the density of states has the following form
\cite{Barnard}:
\begin{equation}
N_d = N_o(E_o-E)^{1/2} \label{eq:dos}
\end{equation}

Where $E_o$ is the energy corresponding to the top of the d-band. Now using
Eq.~\ref{eq:tep},Eq.~\ref{eq:sigma} and Eq.~\ref{eq:tau}, we get
\begin{equation}
S = \left [\frac{\pi ^2 k^2T}{3e}\left (\frac{3}{2E_F}-\frac{dln N_d}{dE}\right )\right ]_{E = E_F} \label{eq:S}
\end{equation}

Differentiating Eq.~\ref{eq:dos} we get
\begin{equation}
\frac{dlnN_d}{dE} = \frac{-1}{2(E_o-E_F)} \label{eq:dN}
\end{equation}

If we substitute this result in Eq.~\ref{eq:tep}, we would get a thermopower which has a negative sign and a linear
temperature dependence . Also $dS/dT$ would have a negative sign, due to the sign of the electronic charge. This
is in contrast to experimental results which show that $dS/dT$ is positive beyond 600$^o$ C in the case of Iron
\cite{Blatt}. To
explain these facts we would have to take into account the fact that in a
ferromagnetic system, the bands are split into spin-up and spin-down bands, due to the magnetic interaction.
Since there are more electrons in the spin-up band compared to the spin-down band the spin-up d band will be filled
to a greater extent than the spin-down band (Fig.1b) \cite{Ziman}. Since the chemical potential (Fermi energy) of
the two-bands, should
be the same, the spin-up band is pushed down with respect to the spin-down band as shown in Fig.1c \cite{Ziman}.
Due to this, the number of available d states at $E_F$ is more in the case of the spin-down band than the spin-up
band. Hence the electrons will be predominantly scattered into the spin-down band.

Taking both the spin-up and spin-down bands into consideration, we can write,
\begin{equation}
N_+ = N_o(E_1-E_F)^{1/2} \label{eq:N+}
\end{equation}
\begin{equation}
N_- = N_o(E_2-E_F)^{1/2} \label{eq:N-}
\end{equation}
 Where $N_+$ and $N_-$ are the density of states at the Fermi surface for the spin-up and the spin-down bands   
respectively and $E_1$ and $E_2$ are the energies corresponding to the top of the spin-up and spin-down bands
respectively (Fig.1c).

Referring to Fig.~1b and Fig.1c, we see that
\begin{equation}
E_1-E_F = E_o-(E_F+E_{\delta}) \label{eq:E1}
\end{equation}
and
\begin{equation}
E_2-E_F = E_o-(E_F-E_{\delta}) \label{eq:E2}
\end{equation}

$E_{\delta}$ is the shift in the Fermi level in the ferromagnetic case with respect to the Fermi energy
in the paramagnetic state.

Taking into account both the spin up and spin-down bands, Eq.~\ref{eq:S} is rewritten as:
\begin{equation}
S = -\frac{\pi ^2k^2T}{3e}\frac{1}{(N_++N_-)}\left [\frac{dN_+}{dE}+\frac{dN_-}{dE}\right ]
\end{equation}

Using Eq.~\ref{eq:N+} and Eq.~\ref{eq:N-} in Eq.~\ref{eq:S}, we have

\begin{eqnarray}
S = \frac{\pi ^2kT}{6e}\frac{1}{[N_o(E_1-E_F)^{1/2}+N_o(E_2-E_F)^{1/2}]} \\ \nonumber
\times \left
[\frac{N_o}{(E_1-E_F)^{1/2}}+ 
\frac{N_o}{E_2-E_F)^{1/2}}\right ]+\frac{\pi ^2k^2T}{2eE_F}
\end{eqnarray}

This can be re-written as
\begin{equation}
S = \frac{\pi ^2k^2T}{6e}\frac{1}{(E_1-E_F)^{1/2}(E_2-E_F)^{1/2}}+\frac{\pi ^2k^2T}{2eE_F}
\end{equation}

Substituting for $E_1$ and $E_2$ we have

\begin{eqnarray}
S = \frac{\pi ^2k^2T}{6e}\frac{1}{[(E_o-E_F)-E_{\delta}]^{1/2}[(E_o-E_F)+E_{\delta}]^{1/2}}+ \\ \nonumber
\frac{\pi
^2k^2T}{2eE_F}
\end{eqnarray}

or
\begin{equation}
S = \frac{\pi^2k^2T}{6e}\left [\frac{1}{(E_o-E_F)^2-E_{\delta}^2}\right ]^{1/2}+\frac{\pi ^2k^2T}{2eE_F}
\label{eq:S1}
\end{equation}

$E_{\delta}$ can be related to the magnetization in the following manner:

The reduced magnetization  is given by:

\begin{equation}
M = \frac{N_+-N_-}{N_++N_-}
\end{equation}

Substituting for $N_+$ and $N_-$ and simplifying the expression, we have

\begin{equation}
M = \frac{-E_{\delta}}{(E_o-E_F)+[(E_o-E_F)^2-E_{\delta}^2]^{1/2}}
\end{equation}
Rearranging the above equation, we obtain the expression for $E_{\delta}$ in terms of $M$.

\begin{equation}
E_{\delta} = -\frac{2M(E_o-E_F)}{1+M^2} \label{eq:Edelta}
\end{equation}

Substituting Eq.~\ref{eq:Edelta} in Eq.~\ref{eq:S1}, we have

\begin{equation}
S = \frac{\pi ^2k^2T}{3e}\frac{1}{(E_o-E_F)}\left [1-\frac{4M^2}{(1+M^2)^2}\right ]^{-1/2}+\frac{\pi ^2k^2T}{2eE_F}
\label{eq:S2}
\end{equation}

The thermopower is
negative at all temperatures, since there is a negative sign from the electronic charge and $E_o>E_F$. This function
is plotted
in Fig.2 for Iron. The values used for the various quantities are: $E_o-E_F = 1.25 eV$ \cite{Mott} and $E_F = 11.1
eV$ \cite{Ashcroft} Iron has a Curie temperature of 770$^o$ C which is seen as a peak in the thermopower. Above
$T_c$ $M=0$ and the contribution to the thermopower is only from the second term in Eq.~\ref{eq:S2}.It is seen that
$dS/dT$
is positive even though $S$ itself is negative. This is in contrast
to the free-electron theory, according to which $S$ and $dS/dT$ have the same sign.

The experimentally measured thermopower for Fe, Co and Ni, fits
this model only above the Debye temperature \cite{Vedernikov}. The maximum seen in the thermopower of Iron
at around 200 K has
been attributed to magnon
drag \cite{Barnard}. However beyond the Debye
temperature all these effects are negligible and diffusion thermopower dominates. The minimum seen in
the thermopower  , maybe due to a combination of phonon drag, magnon drag and diffusion components of the
thermopower. While the diffusion thermopower has a positive slope between $\theta _D$ and $T_c$, the thermopower
due to phonon and
magnon drag might have a negative slope. At low temperatures, the phonon and magnon drag dominates over the
diffusion part, while at higher temperatures the diffusion thermopower is larger in magnitude. Hence $dS/dT$ will
change sign at some temperature near $\theta _D$.

Although Fig.2
matches experimental results \cite{Vedernikov} quite well in the high temperature region, the magnitude is
slightly greater than what
is observed experimentally.   
According to the theory outlined above, the TEP is of the order of -12 $\mu$V/C at 600 K, but experimental results
show that
it is of the order of -5 $\mu$V/C or less.

The present approach clearly predicts the experimentally observed peak in the thermopower near $T_c$

Also it is of
interest
to note that if the mean field value of $M$ is used in the above equations, one obtains a linear variation for $S$,
while if it is assumed that $M \sim ((T_c-T)/T_c)^{0.36}$ according to the Heisenberg model, a non-linear variation
of the
thermopower is obtained.

\section{Conclusions} 

	The Thermopower of ferromagnetic transition metals at high temperatures has been interpreted in terms of a
simple band picture. It has been shown that the calculated results agree qualitatively with experimental
results. Quantitative agreement of the calculated results with experimental results requires
the incorporation of phonon and magnon drag effects. Incorporation of these effects is definitely necessary to
explain the minimum in the thermopower as well as the thermopower at low temperatures.

\vspace{1mm}

{\bf Acknowledgements}

The author wishes to thank Dr.T.G.Ramesh for initiating him into the subject and R.Aravinda Narayanan for useful
discussions and a careful reading of the manuscript.

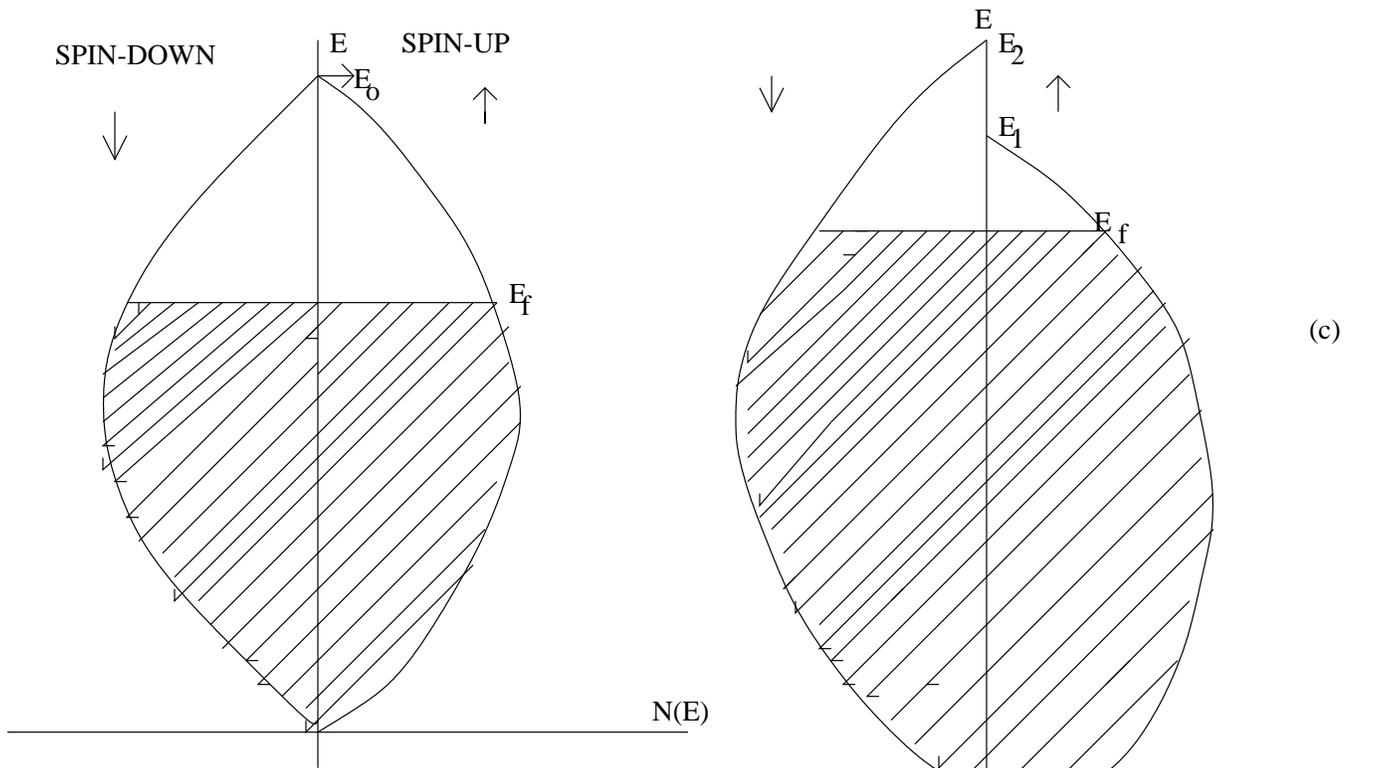
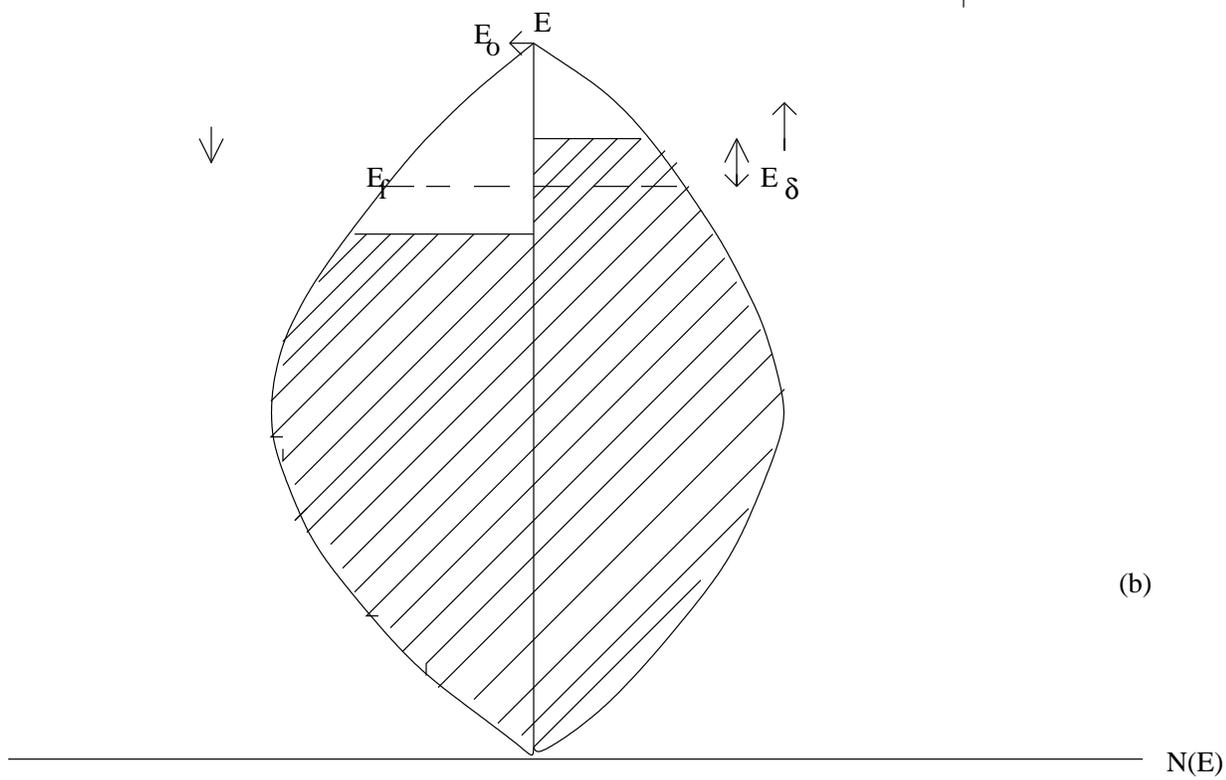
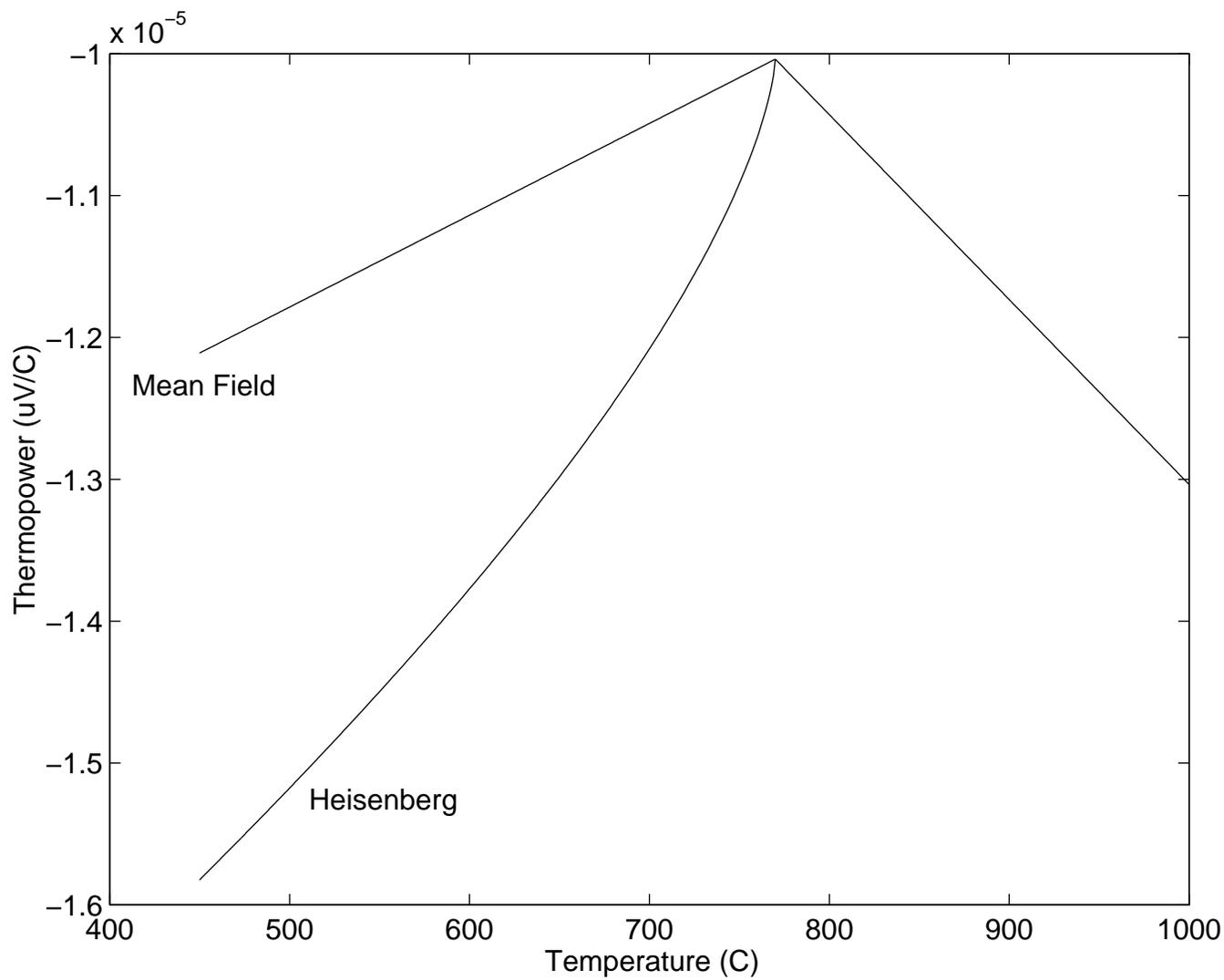
\end{document}